# Zero-Trust Network Access (ZTNA)

Vasilios Mavroudis

Alan Turing Institute
vmavroudis@turing.ac.uk

**Abstract.** Zero-Trust Network Access (ZTNA) marks a significant shift in network security by adopting a "never trust, always verify" approach. This work provides an in-depth analysis of ZTNA, offering a comprehensive framework for understanding its principles, architectures, and applications. We discuss its role in securing modern, complex network environments, which include cloud platforms, Internet of Things (IoT) devices, and hybrid enterprise networks. Our objective is to create a key resource for researchers and practitioners by reviewing critical methodologies, analyzing current implementations, and highlighting open challenges and research directions.

## 1 Introduction

Traditional security models, often referred to as perimeter-based security, operated under the assumption that any user or device within the protected boundary of a network could be trusted [8]. These models rely on firewalls, virtual private networks (VPNs), and demilitarized zones (DMZs) to create a secure perimeter around the network, protecting it from external threats. However, this approach is increasingly inadequate in modern computing environments, where the concept of a fixed perimeter is rapidly disappearing [14,3]. The rise of cloud computing, the proliferation of Internet of Things (IoT) devices, and the expansion of remote workforces have fundamentally altered network topologies, creating more fragmented and complex infrastructures. As a result, perimeter-based security is no longer sufficient, as threats can originate from within the network, devices can operate outside traditional boundaries, and users may require access to resources from multiple locations and platforms [7,13].

Zero-Trust Network Access (ZTNA) emerges as a response to these challenges, offering a more flexible and robust approach to securing modern network environments. The core principle of ZTNA is simple yet powerful: "never trust, always verify". Unlike traditional models that automatically trust any device within the network perimeter, ZTNA assumes that every request for access, regardless of its origin, must be treated with caution and subjected to strict verification. This model shifts the focus from securing the perimeter to securing individual resources [11,13], ensuring that every user and device is authenticated, authorized, and continuously monitored before being granted access to critical network assets.

This work seeks to provide a comprehensive resource on ZTNA, positioning it as a critical approach for modern network security. Beyond detailing its foundational principles and architectures, this work aims to bridge the gap between theory and practice by examining ZTNA's role in addressing real-world security needs across diverse environments, such as cloud platforms, IoT ecosystems, and hybrid enterprise networks. By analyzing ZTNA's applications, challenges, and use cases, we provide a resource for researchers and practitioners interested in deploying adaptive, risk-based security frameworks that can evolve alongside technological advances.

ZTNA operates by dynamically evaluating each access request, considering both the user's identity and the security status of their device [14,2]. This evaluation involves checking factors such as compliance with security policies, device location, and any detected vulnerabilities or malware. Access is granted only when the user and device meet these predefined security criteria. This process reduces risks from insider threats, where malicious or compromised users within the network could gain unauthorized access, as well as from supply chain vulnerabilities, where external vendors or contractors may inadvertently introduce security risks. ZTNA is also effective in mitigating advanced persistent threats (APTs), which often depend on lateral movement within a network after initial access is gained [10].

By enforcing continuous monitoring and adaptive policies, ZTNA enables organizations to implement granular access controls that adjust in response to real-time risk changes. This approach not only strengthens security but also enhances operational efficiency, enabling legitimate users to access needed resources with minimal friction. As organizations increasingly adopt cloud services, remote work models, and IoT devices, ZTNA provides a scalable and adaptable security framework that meets the demands of today's complex, decentralized digital environments [15,1].

The remainder of this paper is organized as follows: Section 2 delves into the foundational principles of ZTNA, including least privilege, continuous authentication, microsegmentation, and policy-based access control, which together form the basis of ZTNA's unique approach to security. Section 3 discusses ZTNA architectures, distinguishing between agent-based and agentless models, and explores Software-Defined Perimeter (SDP) and cloud-based ZTNA implementations. Section 4 addresses the practical challenges organizations face when deploying ZTNA, such as performance overhead, policy complexity, legacy integration, and scalability concerns. Finally, Section 5 presents common ZTNA use cases, including remote work and IoT networks, demonstrating its versatility across diverse environments.

## 2  Key Concepts

ZTNA is underpinned by several foundational principles that shape its approach to network security. These principles are derived from well-established security best practices, which ensure that access to resources is strictly controlled and continuously monitored. By adhering to these principles, ZTNA offers a robust framework that addresses modern security challenges, including insider threats, advanced persistent threats (APTs), and the increased complexity of distributed networks.

### 2.1  Least Privilege

The principle of least privilege is central to ZTNA, ensuring that users are granted only the minimum level of access necessary to perform their tasks [14]. This approach limits the potential damage of a compromised account, as permissions are confined to the user's role, effectively reducing the attack surface. In traditional security models, over-privileged access is a common vulnerability, where users often have broader permissions than necessary. By implementing least privilege, ZTNA helps to contain unauthorized activity within a limited scope, thereby improving overall network resilience. In practice, least privilege is enforced through role-based [12] or attribute-based access controls, which specify permissions based on job functions or other relevant attributes. This targeted access management reduces risks associated with data breaches and internal misuse by ensuring users only interact with the data, applications, and systems relevant to their role.

### 2.2  Continuous Authentication

Continuous authentication departs from traditional one-time authentication methods, which verify identity only at the start of a session. Instead, ZTNA employs adaptive, ongoing authentication processes that continually validate both the identity of users and the security posture of their devices throughout a session [16]. This dynamic approach allows ZTNA to adjust access permissions in real-time based on factors such as user location, device health, and network usage patterns. This continuous validation is crucial in defending against modern security threats, as static credentials, like passwords, can easily be compromised. Techniques such as multi-factor authentication (MFA), biometric verification, and behavioral analytics provide continuous monitoring of user identity and behavior. If suspicious activity is detected, ZTNA can immediately revoke access or prompt additional verification steps, reducing the risk of unauthorized access and mitigating insider threats.

### 2.3  Microsegmentation

Microsegmentation is a core ZTNA concept that divides a network into smaller, isolated zones with individualized security controls [2]. This approach limits attackers' ability to move laterally across the network, containing potential breaches to a single endpoint if successfully compromised. By enforcing strict boundaries between network segments, microsegmentation helps minimize the potential for widespread damage, especially in high-risk environments where sensitive data is involved, such as in healthcare or finance. ZTNA applies granular access controls at the workload level, ensuring that users and devices only access the specific resources they need. Isolating each segment and tightly controlling inter-segment access minimizes exposure, enhancing protection for critical systems against lateral threats.

### 2.4    Policy-Based Access Control

Policy-based access control (PBAC) lies at the heart of ZTNA, governing resource access through predefined security policies [16]. These policies specify who can access which resources, under what conditions, and using which devices [13]. Unlike traditional models with static permissions, PBAC in ZTNA is dynamic and adapts in real-time, leveraging data about users, devices, and network environments to make access decisions. ZTNA policies are customizable and fine-grained, enabling organizations to define specific rules for individual users, groups, or devices. Access can be based on factors such as time of day, geographic location, and device security status. By continuously updating policies across the network, ZTNA ensures that access permissions reflect current security requirements, enabling it to respond effectively to the dynamic nature of modern networks where users access resources from varied locations and devices.

### 2.5    Dynamic Policy Enforcement and Updates

Alongside policy-based access control, ZTNA emphasizes dynamic policy enforcement, where policies are not only applied at initial authentication but also continuously evaluated and adjusted based on real-time conditions [17]. This capability ensures that user permissions remain appropriate as the security context evolves, such as when a device's status changes or a user attempts actions outside their typical behavior [18,1]. Moreover, dynamic policy updates are especially important for countering threats such as APTs, where malicious actors gain extended access to a network and gradually increase their privileges. By continually adjusting access controls, ZTNA prevents such attackers from exploiting static permissions, significantly reducing their ability to escalate privileges or move laterally within the network.

## 3    Architectures and Implementation

ZTNA architectures are typically divided into two main categories: Agent-Based ZTNA and Agentless ZTNA, each tailored to different security needs and operational contexts. To address specific network requirements and deployment environments, these approaches can incorporate specialized implementations, such as Software-Defined Perimeter (SDP) and Cloud-Based ZTNA. Both architectures rely on dynamic, risk-based access decisions to maintain continuous security, adjusting permissions in real time based on evolving risk assessments.

### 3.1  Agent-Based ZTNA

Agent-Based ZTNA involves installing a software agent on each device requiring access to the network [16]. This agent enforces security policies locally and continuously monitors device health and compliance, allowing for fine-grained security management based on device-specific characteristics (e.g., OS version, patch level, configuration). While this approach provides high control, managing agents across a large range of devices can pose challenges, especially in BYOD environments. However, agent-based ZTNA is well-suited for managed environments with centralized device administration, offering granular access control and dynamic compliance adjustments.

### 3.2  Agentless ZTNA

Agentless ZTNA, in contrast, leverages network infrastructure components—such as secure web gateways, proxies, and identity-aware firewalls—to enforce security policies without requiring software installation on devices. This model is especially effective in environments with heterogenous device types (e.g., low-powered edge devices, where installing agents may not be feasible. Although agentless ZTNA lacks the device-level granularity of agent-based solutions, it offers greater ease of deployment and scalability. This approach is ideal for organizations that prioritize broad compatibility and flexibility but may have limitations in continuously monitoring device health, which is more readily accessible with on-device agents.

### 3.3  Specialized Implementations

ZTNA architectures often incorporate additional models to meet specific network and security needs. Two prominent implementations are the Software-Defined Perimeter (SDP) and Cloud-Based ZTNA.

**Software-Defined Perimeter**  The Software-Defined Perimeter (SDP) model enhances security by concealing network resources from unauthorized users and granting access only to authenticated entities. SDP leverages a control plane to authenticate users and devices, reducing lateral movement risks by segmenting access [9]. SDP is composed of key components: the Initiating Host, Policy Decision Point (PDP), and Policy Enforcement Point (PEP). The Initiating Host authenticates with the PDP, which verifies identity and compliance before granting access. The PEP then enforces the PDP's access decision by acting as the gatekeeper, allowing or denying connections based on the authorized parameters. This identity- and context-driven security framework makes SDP highly adaptable to both hybrid cloud and on-premises environments. Note that SDP components may vary depending on the use case; for instance, [6] introduces additional elements to enable real-time access adjustments, particularly suited for environments with high device variability.

**Cloud-Based ZTNA** Cloud-Based ZTNA solutions are designed to address the unique security challenges of cloud environments, including multi-tenancy and dynamic resource allocation [13]. Unlike traditional on-premises security models, cloud-native ZTNA enforces layered access controls—from device and network entry points to the application layer—ensuring security coverage in cloud-first architectures. Cloud-based ZTNA continuously evaluates access based on user identity, device health, and behavior patterns, adapting as conditions change. This dynamic approach is essential in cloud environments with distributed resources accessed from multiple locations [11]. It is especially beneficial in hybrid or multi-cloud infrastructures, providing consistent security policies across providers.

## 4  Deployment Challenges

While ZTNA offers considerable advantages in securing modern network environments, several challenges complicate its deployment and effectiveness. As organizations move towards zero-trust models, they must navigate issues related to performance, policy complexity, integration, and scalability.

### 4.1  Performance Overhead

One of the primary challenges of ZTNA is the performance overhead associated with continuous verification. Unlike traditional security models, which authenticate users and devices only once per session, ZTNA involves ongoing assessments to verify the legitimacy of each access request [5]. This constant verification process can introduce latency, particularly in high-traffic environments where hundreds or thousands of users and devices request access simultaneously. For organizations with high performance requirements, such as financial services or real-time data processing, managing this performance impact without compromising security can be complex [18].

### 4.2  Complexity in Policy Management

ZTNA's granular access control, while it provides enhanced security, also introduces complexity in policy management. Defining and implementing access policies in a ZTNA model requires careful consideration of multiple factors, including user roles, device types, geographic location, and behavioral context [12]. Organizations need to establish policies that are both specific enough to be effective and flexible enough to account for changes in user behavior and network conditions. This complexity increases the administrative burden, especially in large organizations where security teams may struggle to manage a vast number of unique access policies effectively.

### 4.3  Integration with Legacy Systems

Integrating ZTNA with legacy systems presents another major challenge, as many organizations continue to rely on older systems that may not fully support ZTNA's rigorous access controls. Legacy infrastructure often lacks the compatibility needed to meet ZTNA requirements, such as continuous authentication and microsegmentation [14]. For organizations that depend on these systems, retrofitting ZTNA can be difficult, requiring additional resources, custom solutions, or even a complete overhaul of existing network architecture. Without integration, however, these legacy systems can become weak points within the broader security framework.

### 4.4  Scalability

Scalability is a critical consideration in ZTNA deployment, particularly in large-scale networks with diverse devices and users [14]. Implementing ZTNA across a vast, heterogeneous environment can strain network resources and present challenges in managing consistent policies across various devices. As networks expand to include cloud-based resources, remote users, and IoT devices, maintaining uniform access controls becomes increasingly complex [13]. Scalability issues can lead to gaps in security if ZTNA controls are inconsistently applied, particularly in multi-cloud or hybrid cloud environments where resources are distributed.

## 5  Use Cases

The design principles underlying ZTNA enable its application across a range of complex network environments, each with distinct security requirements. This section explores specific use cases where ZTNA's continuous authentication and strict access control provide meaningful advantages.

### 5.1  Remote Work

The increase in remote work has created a need for secure access to corporate resources from beyond traditional office boundaries [6]. ZTNA addresses this by enforcing strict identity verification and dynamic access controls, allowing remote employees to securely access necessary resources, even from personal devices or unsecured networks. This approach is essential for maintaining security in a distributed workforce model and supports organizations adapting to remote work environments.

### 5.2  Bring Your Own Device

BYOD environments introduce unique security challenges, as personal devices often lack the security protocols of corporate-managed devices. ZTNA mitigates these risks by applying secure access controls that protect network resources without infringing on user privacy [1]. This enables employees to use their personal devices while reducing risks associated with unmanaged or insecure endpoints.

### 5.3  IoT and Edge Device Security

Edge and IoT devices are particularly vulnerable to cyberattacks due to their often limited security capabilities. ZTNA enhances the security of these devices by enforcing strict access control and continuous authentication, dynamically assessing the risk associated with each device and applying granular access permissions. This approach is especially valuable for IoT deployments in sectors such as healthcare and industrial environments [4], where compromised edge or IoT devices could lead to significant operational disruptions or privacy risks [11].

### 5.4  Next-Generation Networks

As network technology advances, future environments will likely encompass large, diverse ecosystems of connected devices, including IoT, autonomous systems, and critical infrastructure. These next-generation networks require adaptable security models to address high device diversity, large scale, and real-time demands. ZTNA's continuous verification of each access request—at both the user and device levels—offers a scalable, granular control approach that secures access regardless of location or device type. With its principles of least privilege and continuous authentication, ZTNA is well-positioned to support future networks where flexibility and real-time responsiveness are essential. Further research will be needed to evaluate ZTNA's scalability and performance as network technologies continue to evolve.

## 6  Conclusion

Zero-Trust Network Access represents a pivotal shift in network security, overcoming the fundamental limitations of traditional perimeter-based models. By enforcing continuous identity verification and strict, policy-driven access controls, ZTNA mitigates risks posed by insider threats, compromised devices, and evolving attack vectors. This approach transforms security from a static, boundary-focused model to a dynamic framework that aligns access permissions with real-time risk. As networks grow more complex and dispersed with cloud, IoT, and remote access technologies, ZTNA's principles of least privilege, microsegmentation, and adaptive policy enforcement provide a robust defense, particularly against the lateral movement of attackers. This adaptability across diverse environments—from remote work to IoT—positions ZTNA as a vital component of resilient, future-proof security strategies, aligning with the increasingly decentralized nature of digital ecosystems.